# CLTree: A Tool for Annotating, Rooting, and Evaluating Phylogenetic Trees Leveraging Genomic Lineages

Guanghong Zuo[1,a,*]

[1]*Wenzhou institute, University of Chinese Academy of Sciences, Wenzhou, Zhejiang 325001, China*

[*]Corresponding author E-mail: ghzuo@ucas.ac.cn (Zuo GH).
[a] ORCID: 0000-0002-7822-5969.

**ABSTRACT**: Collapse Lineage Tree (CLTree) is a software tool that annotates, roots, and evaluates phylogenetic trees by using lineages. A recursive algorithm was designed to annotate the branches by the common taxonomic lineage of its descendants in a rooted tree. For an unrooted tree, it determines the root that best conforms to the taxonomic system based on the aforementioned lineage annotations. Based on the lineage annotations of notes, CLTree infers the monophyly of taxonomic units and quantifies the concordance between the phylogenetic tree and the taxonomic system base on Shannon entropy. The core algorithm of CLTree is highly efficient with linear complexity, capable of processing phylogenetic trees with 17,955 terminal nodes within one second. We believe that CLTree will serve as a powerful tool for study of evolution and taxonomy.

# Introduction

The phylogenetic tree serves as a fundamental tool for depicting the evolution of species and genes[1–3]. Phylogenetic trees are typically constructed using genetic information, such as DNA sequences and amino acid sequences, to infer evolutionary relationships[4–6]. Several approaches are available for constructing phylogenetic trees from gene sequences, such as maximum likelihood[7,8], Bayesian inference[9–11], and distance-based methods[12–14]. By analyzing the genetic characteristics and evolutionary pathways of species, it can address challenges in species delimitation[15]. Phylogenetic trees are instrumental in evaluating the impact of species extinction on evolutionary dynamics and in forecasting shifts in species distribution patterns. For already constructed phylogenetic trees, two issues remain: first, how to evaluate the quality of the phylogenetic tree[16]; second, many methods generate only unrooted trees, which indicate relative relatedness rather than evolutionary relationships among species, thus requiring rooting[17].

The prevalent approach for evaluating phylogenetic trees is the bootstrap method[18–21], which assesses branch stability by calculating data consistency through bootstrap sampling of aligned sequences. It is crucial to highlight that bootstrap methods, frequently employed in phylogenetic analysis, mainly assess the consistency of data rather than the definitive accuracy of evolutionary relationships. Regarding rooting, commonly used methods include the outgroup method [22] and branch length distribution methods[23] such as midpoint rooting[24], minimum variance rooting[25], and minimum ancestor deviation[26]. The outgroup method roots the tree based on the sequences close but outside the study group, but a qualified outgroup is not available in some case, e.g. the all-living species tree. The branch length distribution methods determine the optimal root based solely on statistical analysis of branch length and genomes distribution bias can also disrupt the final choice of roots. Biological evolution and taxonomy are two sides of a coin[27–29]; leveraging taxonomic information in conjunction with statistical methods for phylogenetic tree evaluation and rooting provides a novel perspective for investigating evolutionary relationships[30].

In this work, we introduce our recently developed computational tool, the Collapsed Lineage Tree (CLTree). We have developed a recursive algorithm that leverages the tree structure of phylogenetic trees, starting from the root node and progressively annotating each node with the common taxonomic lineage of its descendant nodes. Subsequently, it calculates the concordance of the phylogenetic tree with lineages at every taxonomic rank based on these annotated labels, thereby evaluating the phylogenetic tree. If the phylogenetic tree is unrooted, the algorithm also integrates the lineage labels with branch lengths to infer the root. Algorithmic analysis suggests that the software's core algorithm operates linearly, allowing for efficient processing of large datasets, a feature that is crucial in phylogenetic tree construction and analysis.

## Algorithms and Implementations

**Taxonomy input for Tree**

Figure 1 illustrates the workflow of the CLTree software. The majority of the workflow is dedicated to determining the lineage of all sequences, i.e. the terminal nodes (denoted as leaves) of the phylogenetic tree. It should be noted that taxonomy is dynamic, and as there are several available sources for it, the lineage of a leaf may not be obvious. In our study, we utilize the dump files from NCBI Taxonomy database[31] as the basic information source, because most of genomes of phylogenetic trees have the corresponded TaxID of the NCBI Taxonomy database. And we also offer user-friendly interfaces for manual lineage updates based on other taxonomy sources, such as Bergey's Manual[32,33], LPSN[34,35], and relevant published papers. In CLTree, we also provide a script that obtains LPSN lineage based on the genus.

**Annotate Phylogenetic Tree by Lineage**

Once the lineage of the leaves is obtained, the core module of CLTree, known as "Collapse and Rooting", is executed. For rooted trees, we have designed a depth-first recursive algorithm that commences from the root of the phylogenetic tree (as depicted in Algorithm 1). Generally, for a binary tree with N leaves, a rooted tree has N-1 internal nodes (denoted as branches), and the number of leaves of a branch is incrementally distributed. If each branch is queried and compared individually, the algorithm complexity would be $O(N^3)$. However, our algorithm reduces the complexity to linear by examining the nodes according to the tree structure.

For unrooted trees, it is necessary to determine the root and transform them into rooted trees before comparing them with taxonomy. An unrooted binary tree has *2N-3* nodes, and theoretically, there are *2N-3* candidates for the root. Our objective is to find the root that best fits the taxonomy. Our research indicates that the following operations can contribute to achieving this goal. Initially, we randomly select a root and carry out the aforementioned annotation process. Subsequently, we employ the method outlined in Algorithm 2 to identify a candidate root, and then update the lineage annotation. It is important to note that there may be several root candidates that best fits the taxonomy. To ascertain the final root, further adjusting based on branch lengths is required. In CLTree, we also implemented various branch length rooting methods, e.g., minimal variance, minimal ancestor deviation, midpoint, pairwise midpoint root, and minimal depth. For algorithmic details about these options please refer to the manual.

**Measuring Diversity via Shannon Entropy**

All leaves in the phylogenetic tree are classified into distinct groups at each taxonomic rank, where each leaf belongs to exactly one group, representing a classic classification problem. According to information theory[36], the Shannon entropy at a taxonomic rank is defined as:

$$H_{taxa} = -\sum_i \frac{n_i}{N} log_2 \frac{n_i}{N}$$

where $n_i$ with $\sum n_i = N$ is the number of leaves in a group (taxonomic unit), and N is the total number of leaves in the phylogenetic tree. To obtain the Shannon entropy of the groups at a taxonomic rank of a rooted annotated tree, we identified the clades for the taxonomic units on the phylogenetic tree. In a phylogenetic tree, a node is considered a clade when the taxonomic group rank of the branch is lower than that of its parent branch, indicating a group of organisms that share a common ancestor. Thus a clade is correspond only one taxonomic unit at a taxonomic rank, while a taxonomic unit may correspond to more than one clades. It is obvious that the clades at a taxonomic rank also divide the leaves into groups, with each leaf belonging to only one group. Thus, the Shannon entropy of the clades at a taxonomic rank is defined as:

$$H_{tree} = -\sum_i \frac{m_i}{N} log_2 \frac{m_i}{N}$$

where $m_i$ with $\sum m_i = N$ is the number of leaves on a branch. The maximum value of the Shannon entropy for clades is: $H_{max} = -\sum_i \frac{1}{N} log_2 \frac{1}{N}$ which corresponds to the scenario where each leaf belongs to a different taxonomic unit from its siblings. It is evident that $H_{taxa} \leq H_{tree} \leq H_{max}$. To make the final metric more comprehensible, the output values in the CLTree system are scaled, and the degree of consistency between the phylogenetic tree and taxonomy at a taxonomic rank is measured by the entropy reduction ratio, expressed as:

$$\widetilde{\Delta H} = \frac{H_{max} - H_{tree}}{H_{max} - H_{taxa}}$$

where $0 \leq \widetilde{\Delta H} \leq 1$. A value of 0 represents the worst case, where each leaf belongs to a different taxonomic unit from its siblings; a value of 1 represents the best case, where all taxonomic units at this rank are monophyletic on the phylogenetic tree.

**Implementation**

The CLTree software is developed using standard C++ and follows an object-oriented design. The source code of CLTree is managed using CMake. And it can be compiled and executed on Linux/Unix, Macintosh, and Windows platforms. The "cltree" is the main command of the software. It can either execute the entire process from scratch in a single run or utilize subcommands to control the execution of specific tasks. For more options regarding CLTree and its subcommands, please refer to the manual. The CLTree software, including example data, documentation, and source codes, are freely available for academic users.

**Examples**

To illustrate the algorithm's effectiveness, we utilized the CVTree method [37,38] to construct a phylogenetic tree using 16S rRNA data from the All-Species Living Tree project[39,40] . There are 17,955 leaves on this tree after excluding sequences that were excessively long or short. CLTree handled this 17,955-leaf phylogenetic tree in a second on my laptop. The comparison results were presented from two perspectives: taxonomy and phylogenetic tree. Table 1 presents a summary of phylogenetic analysis results from the CLTree, including the count of monophyletic taxon units, which correspond to unique branches on the phylogenetic tree, the Shannon entropy values for both the taxonomic and phylogenetic structures, and the ratio of entropy reduction. Besides the summary results, the CLTree command also outputs detailed results (see supporting information 2).

Figure 2 presents the results from the perspective of a phylogenetic tree. To enhance the visual interpretation of phylogenetic relationships, the CLTree command can output not only a Newick-formatted tree with annotations but also files suitable for display on the iToL website [41,42] (https://itol.embl.de/). These files include the Newick tree and scripts for labeling, coloring, and interactive display on the iToL platform. As demonstrated in Figure 2, CLTree effectively positions the root of the tree at the divergence point between bacteria and archaea. The phylogenetic tree encompasses two of the three domains, with the third domain of eukaryotes having evolved from the archaeal domain[43–45]. Essentially, there are few sequences from any species that lie outside of this phylogenetic tree, making it challenging to identify a suitable outgroup for the tree. Moreover, the vast disparity in the number of sequences between bacteria and archaea often leads to errors when employing branch length statistics to process such datasets. However, the CLTree algorithm offers an excellent solution to the rooting problem under these circumstances.

Figure 2 also visually illustrates the distribution of various phyla within the phylogenetic tree, where six phyla are represented by only one sequence, naturally corresponding to a branch on the tree; thirty phyla, which contain multiple sequences, also cluster together into single branches on the tree; however, there remain five phyla: Methanobacteriota, Mycoplasmatota, Pseudomonadota, Bacillota, and Chloroflexota, which are distributed across multiple branches on the phylogenetic tree, failing to form a single cohesive branch. This indicates inconsistencies in the classification and phylogenetic relationships of these particular taxa [38].

## Conclusion

CLTree annotates phylogenetic trees with lineage, subsequently identifying the root of the phylogenetic tree based on the annotations of nodes, and assessing the consistency between phylogenetic tree and taxonomy. The enhanced CLTree algorithm, with its linear complexity, ensures greater efficiency and the ability to process phylogenetic trees with tens of thousands of leaves. It presents comparative results from two distinct perspectives: taxonomy and phylogenetic tree. We believe that CLTree will serve as a powerful tool for the study of evolution and taxonomy. It is worth noting that, since taxonomy also follows a tree structure, this algorithm can be adapted to assess the similarity between two phylogenetic trees.

## Code availability

The CLTree software code is available for download at the following links: https://github.com/ghzuo/collapse and https://gitee.com/ghzuo/collapse

## Author Contributions

GZ designed the study, wrote the source code, performed the analysis, and wrote the manuscript.

## Competing interests

The authors have declared that no competing interests exist.

## Acknowledgments

GZ thanks the Wenzhou institute, University of Chinese Academy of Sciences (Grant No. WIUCASQD2021042).

# Pseudocode for Algorithms

---

**Algorithm 1** Collapse Algorithm for Rooted Tree
---
**Require:** Root Node of Unannotated Tree
**Ensure:** Root Node of Annotated Tree
 1: Initialization: get lineages of leaves
 2: **for** $node \in root.children$ **do**
 3:     ANNOTATE(node)
 4:
 5: **function** ANNOTATE($node$)
 6:     **if** $nd.children.size \neq 0$ **then**
 7:         **for** $nd \in node.children$ **do**
 8:             ANNOTATE(nd)
 9:         $node.lineage \leftarrow$ common lineage of $note.children$

---

**Algorithm 2** Find Root Candidates for Unrooted Tree
---
**Require:** Pre-annotated Tree
**Ensure:** Root Candidates
 1: Initialization: $candidates$
 2: $topRank \leftarrow$ the rank of root
 3: EXAMINE($root, topRank, candidates$)
 4:
 5: **function** EXAMINE($node, topRank, candidates$)
 6:     **if** $\cap rank(node.children) < topRank$ **then**
 7:         add $node$ to $Candidates$
 8:     **else**
 9:         **for** $nd \in node.children$ **do**
10:             **if** $rank(nd) = topRank$ **then**
11:                 EXAMINE($nd, topRank, candidates$)

**Table 1** The number of monophyletic taxon units with multiple genomes on the phylogenetic trees. The number in parentheses in the first column indicates the number of taxonomic units that contain unique genome and multiple genomes.

| Taxon Rank | $N_{solo}$ | $N_{multi}$ | $H_{tree}$ | $N_{taxa}$ | $H_{taxa}$ | $\widetilde{\Delta H}$ |
|---|---|---|---|---|---|---|
| Domain | 0 | 2 | 0.208 | 2 | 0.208 | 1.000 |
| Phylum | 6 | 30 | 2.707 | 41 | 2.506 | 0.983 |
| Class | 21 | 59 | 4.997 | 97 | 3.781 | 0.882 |
| Order | 52 | 121 | 7.142 | 231 | 5.678 | 0.827 |
| Family | 133 | 270 | 8.545 | 589 | 7.294 | 0.817 |
| Genus | 1657 | 1220 | 10.907 | 3436 | 9.968 | 0.774 |
| Species | 17476 | 97 | 14.109 | 17652 | 14.098 | 0.601 |

## Figure Legends

**Figure 1 Flowchart of CLTree for Lineage.**

The workflow of the CLTree command was to start from the green ellipse, follow the arrow line, normal end at the blue box. In the workflow, the cltree command automatically checks the cache data of every step to avoid redundant calculation.

**Figure 2 Collapsed Tree at Phylum Rank.**

In this phylogenetic tree, branches belonging to the same phylum are collapsed into a triangle and are color-coded according to their phylum, corresponding to the surrounding color bars. The phylum names are marked in blue or red, depending on whether they belong to Bacteria or Archaea, respectively.

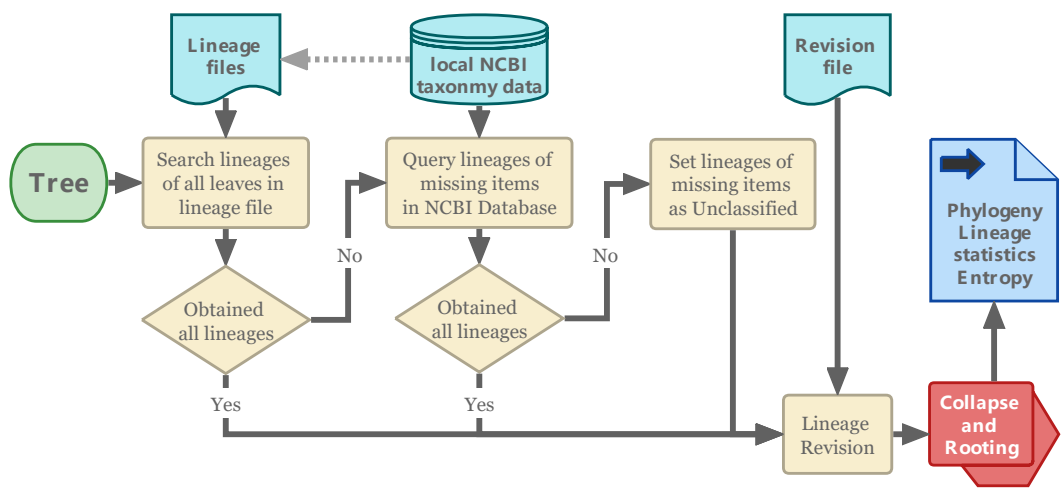

**Figure 1 Flowchart of CLTree for Lineage**

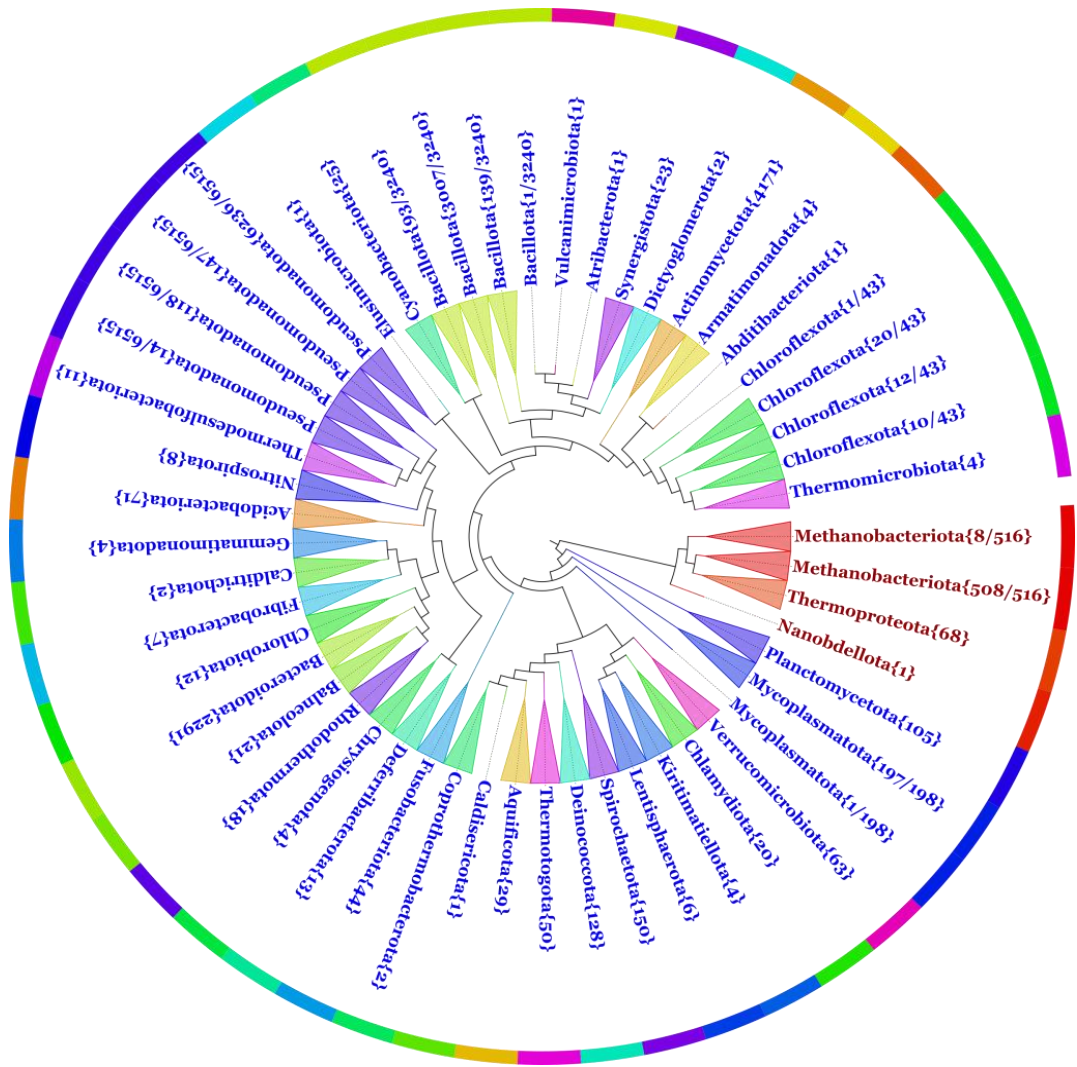

**Figure 2 Collapsed Tree at Phylum Rank.**